# Empirical Predictions for the Period Distribution of Planets to be Discovered by the Transiting Exoplanet Survey Satellite

Jonathan H. Jiang[1], Xuan Ji[2], Nicolas Cowan[3], Renyu Hu[1], Zonghong Zhu[2]

1.  Jet Propulsion Laboratory, California Institute of Technology, Pasadena, California, U.S.A.
2.  Department of Astronomy, Beijing Normal University, Beijing, China
3.  Department of Earth and Planetary Sciences and Department of Physics, McGill University, Montreal, Canada

**Abstract**

Launched in April 2018, NASA's Transiting Exoplanet Survey Satellite (TESS) has been performing a wide-field survey for exoplanets orbiting bright stars with a goal of producing a rich database for follow-on studies. Here we present estimates of the detected exoplanet orbital periods in the 2-minute cadence mode during the TESS mission. For a two-transit detection criterion, the expected mean value of the most frequently detected orbital period is 5.01 days with the most frequently detected range of 2.12 to 11.82 days in the region with observation of 27 days. Near the poles where the observational duration is 351 days, the expected mean orbital period is 10.93 days with the most frequently detected range being from 3.35 to 35.65 days. For one-transit, the most frequently detected orbital period is 8.17 days in the region with observation of 27 days and 11.25 days near the poles. For the entire TESS mission containing several sectors, we estimate that the mean value of orbital period is 8.47 days for two-transit and 10.09 days for one-transit, respectively. If TESS yields a planet population substantially different from what's predicted here, the underlying planet occurrence rates are likely different between the stellar sample probed by TESS and that by Kepler.

## 1. Introduction

The Transiting Exoplanet Survey Satellite (TESS), a NASA Explorer-class exoplanet finder mission, has been in orbit since 18 April 2018. TESS is monitoring 26 observational sectors, each covering about 2300 square degrees of the sky [*Ricker et al.* 2015; *Barclay et al.* 2018]. For most of TESS's field-of-view (FOV), the spacecraft will observe the selected stars in the TESS Input Catalog (TIC) to measure the brightness at a 2-minute cadence for 27 days. It will also obtain 30-minute cadence observations of all objects in the TESS fields of view, but we do not discuss the yield of this kind of observation in this paper. Over the mission design lifetime of 2 years, TESS will continuously survey ~ 85% of the sky for 27 days. Thus, the majority of TESS sensitivity will be devoted to short-period exoplanets orbiting closer to their parent stars. However, the survey methodology has been designed such that certain parts of the sky will be surveyed across multiple runs with overlapping FOVs, enabling longer duration of observations, especially at the ecliptic poles [*Ricker et al.* 2015]. Therefore, TESS will have additional observational durations of 54, 81, 108, 189 and up to 351 days in different regions of the celestial sphere to allow for sensitivity to exoplanets with different orbital periods. The region surrounding the ecliptic pole will be especially visible in multiple TESS FOVs with observational baselines of close to a year, enabling the search for planets in Earth-like orbits.

To confirm the exoplanets detected by TESS and to determine their orbital period, at least two transits are required. Simulations by *Sullivan et al.* [2015], *Bouma et al.* [2018], *Huang et al.* [2018] and *Ballard* [2018] have predicted the exoplanet properties yielded by TESS's two or more transit events. *Sullivan et al.* [2015] focused on the model for the relevant stellar and planetary populations in order to forecast the properties of the brightest transiting planet systems in the sky. These studies used a synthetic stellar population rather than a real catalog. *Barclay et al.* [2018] updated the simulation of TESS's yield using the TESS Input Catalog (TIC) Candidate Target List (CTL); their detection model is similar to *Sullivan et al.* [2015]. *Huang et al.* [2018] improved the simulation using empirically-based simulation of multi-planet systems, and updating the photometric noise model and so on. All of their simulations used the planet occurrence rates from *Fressin et al.* [2013] for stars with effective temperature ($T_{eff}$) $\geqslant$ 4000K, and from *Dressing and Charbonneau* [2015] for stars with $T_{eff}$ < 4000K. *Fressin et al.* [2013] determined the rate of occurrence of planets assuming initial distributions of planets as a function of planet size and orbital period provided by *Howard et al.* [2012]. And the true planet population was obtained by simulating it in detail with no prior assumption and then adding their simulated false positives to match the list of *Batalha et al.* [2013] which only contain the planets detected by Kepler during the first 16 months after accounting for the detectability of both planets and false positives. Now we have the full sample of 4-years of observation which has less bias on orbital period. *Dressing and Charbonneau* [2015] used the Kepler data set, but they still have some dead zones at orbital periods of 60 to100 days and radii of 0.5 to 1 solar radius ($R_{\oplus}$). The occurrence rates from both *Fressin et al.* [2013] and *Dressing and Charbonneau* [2015] are based on Kepler data, but are limited in orbital period up to 85 and 200 days, respectively, beyond which the statistics are still too poor to provide results.

The possible yield of TESS's single transiting planets has also been studied by *Villanueva et al.* [2018], in which they predict up to 241 single-transit events during TESS's primary mission.

*Villanueva et al.* [2018]'s planet occurrence rates were also based on the result of *Fressin et al.* [2013] and *Dressing and Charbonneau* [2015]. These planet occurrence rates are only complete to periods of ∼ 100 days, but they extrapolate these rates to periods of > 1000 days to explore the probability of finding planets at longer periods. They assume the occurrence of the longer orbital period is equal to that of the longest complete orbital period. Note that the sample of orbital period within 503 days has a completeness > 90% according to our following analysis. We can dependably predict the occurrence of the single transit of which the orbital period is within 503 days to check the result.

To summarize, previous TESS yield predictions are mostly based on simulated data informed by Kepler yields. In order to obtain the actual occurrence rate of exoplanets, it is necessary to correct the data for detection bias and incompleteness and it always leads to the result limited in a small range due to the incompleteness of data. As TESS uses the same detection method as Kepler, both should be subject to the same type of detection bias. Since the Kepler mission provided biased data for us, we can perform our analysis directly on these data. All we need to do is to compare the difference of target stars and the properties of the detector between two missions. Then the distribution of TESS's transiting planets can be derived from those of Kepler, through which we can prevent the uncertainties of estimating the occurrence of the planet with a given star and eliminate the extra steps. And the result will not be limited in such a small range like before.

In this study, we compute the probabilities of exoplanet orbital period distributions detected in each of TESS's observational duration regions assuming a two-transits detection criterion and one-transit detection criterion. The probability density function of orbital period is calculated based on the observed data during the Kepler mission. We first review the detection bias of different exoplanet detection methods (Section 2), followed by a description of the methodology for computing the orbital period distributions of detected exoplanets (Section 3). The results, uncertainty analysis, and comparisons with previous studies are presented in Section 4, followed by a summary and discussion in Section 5.

## 2. The Detection Bias

The orbital period of exoplanets is critical for understanding the formation and evolution of the star system that hosts the exoplanet [*Wang and Peng* 2015]. It also provides important knowledge about the orbital radius, which plays a key role in determining whether or not the planet can support life [*Kopparapu et al.* 2013].

Using data from the NASA Exoplanet Archive (https://exoplanetarchive.ipac.caltech.edu/; "Archive", here after), we plot in Figure 1 the normalized histogram of exoplanets' orbital periods (P) detected by different methods: transit, radial velocity (RV) and others. The exoplanets detected using the transit method are almost all from the Kepler mission. As shown in Figure 1, the distribution of exoplanet orbital periods varies from the detection methods. The most detectable orbital periods using transit photometry range from 1 to 100 days, while for the RV method the most detectable are exoplanets with orbital periods of 200 to 2000 days. A combination of all other methods yields exoplanets with a wide range of orbital periods from a few days to a few tens of years.

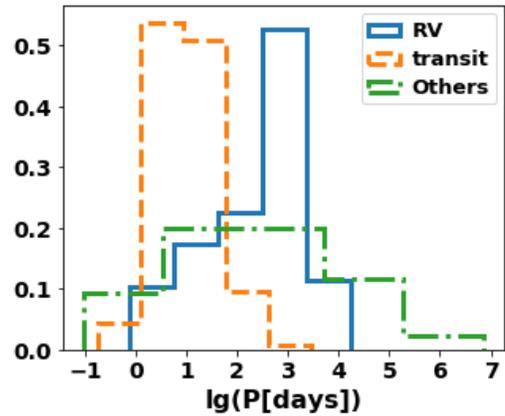

**Figure 1:** Normalized histogram of exoplanets' orbital periods (P) detected by different methods: transit, radial velocity (RV) and others. The x-axis is base 10 logarithmic T (days) and the y-axis is the probability distribution of the orbital periods normalized by the total number of samples.

The method of detection of both transit and RV detected exoplanets depends on the exoplanets' orbital period. We plotted the Hertzsprung-Russell (HR) diagram for their host stars, as shown in Figure 2, where the x-axis shows the effective temperature of the host star in Kelvin $T_{eff}$(K) provided in the Archive. The y-axis shows the log base 10 luminosity ($L$) divided by solar luminosity ($L_\odot$), which is derived using the Stefan-Boltzmann relation $L/L_\odot = (R_s/R_\odot)^2 (T_{eff}/T_\odot)^4$, where $R_s$ is the radius of the host star, also available in the Archive, and $R_\odot$ is the solar radius. It can be seen that transit method detects mainly exoplanets orbiting the main-sequence stars, while almost all exoplanets orbiting red giants are detected by RV [e.g. *Jiang and Zhu* 2018]. This means we should reject the hypothesis that the two sets of data, transit and RV, come from the same distributions.

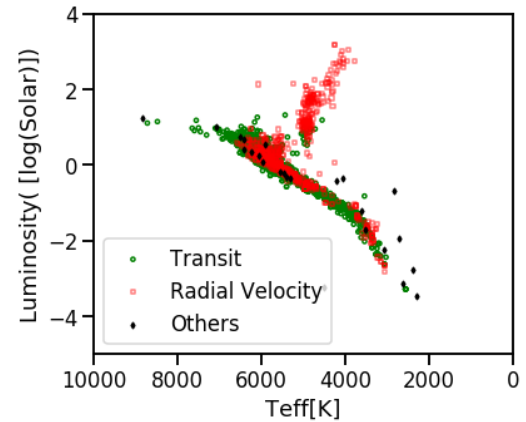

**Figure 2:** The HR diagram of exoplanet host stars. The green circles are stars of which planets are detected by the transit method, the red open squares are stars with planets be detected by RV method, and the black diamonds are stars with planets be detected by other methods. The gap in the middle of the main sequence for the red squares is likely due to the selection effect of radial velocity.

The transit method is so far the most effective and sensitive method for detecting extrasolar planets, particularly from space telescopes. The limitation and biases of this method have been studied in details by *Kipping and Sandford* [2016]. To the first order, assuming the stellar disc is of uniform brightness and neglecting any flux from the planet, the ratio of the observed relative change in flux can be approximated as $\Delta F/F = R_p^2/R_s^2$, where $R_p$ is the radius of the exoplanet. The transit may be too faint to distinguish if $R_p/R_s$ is too small. A planet's transit lasts only a tiny fraction of its total orbital period, so that even when we observe a star with a transiting planet, it is unlikely to repeat the observation if the telescope observes the system for less than the orbital period of the planet. Even if the transit repeats, we are less likely to detect the outer exoplanets with longer orbital period compared to the inner one with the same radius due to the lower signal-to-noise, which is related to the time of transits.

Figure 3a shows $R_p/R_s$ versus orbital period using data from the Archive. For comparison, Figure 3b shows the $M_p/M_s$ versus orbital period, where $M_p$ and $M_s$ are the radii of the exoplanet and host star, respectively. These distribution results in Figure 3a,b are similar to the simulated distribution by *Sullivan et al.* [2015], except that Kepler's transiting exoplanets seem to be divided into two groups above and below the $R_p/R_s \sim 5$ in Figure 3a, which is a topic of ongoing investigation [e.g. *Fulton et al.* 2017]. Also noted is that the mass data for the transiting exoplanets are limited to those with a short period (< 5 days), only a small proportion of RV exoplanets have such a short period, which is in contradiction with the fact that the RV method is more sensitive to the closer exoplanet if $M_p/M_s$ ratios are equal. Due to the strong selection bias shown in Figure 3, it is difficult to obtain the accurate occurrence of the planet as a function of the planetary properties.

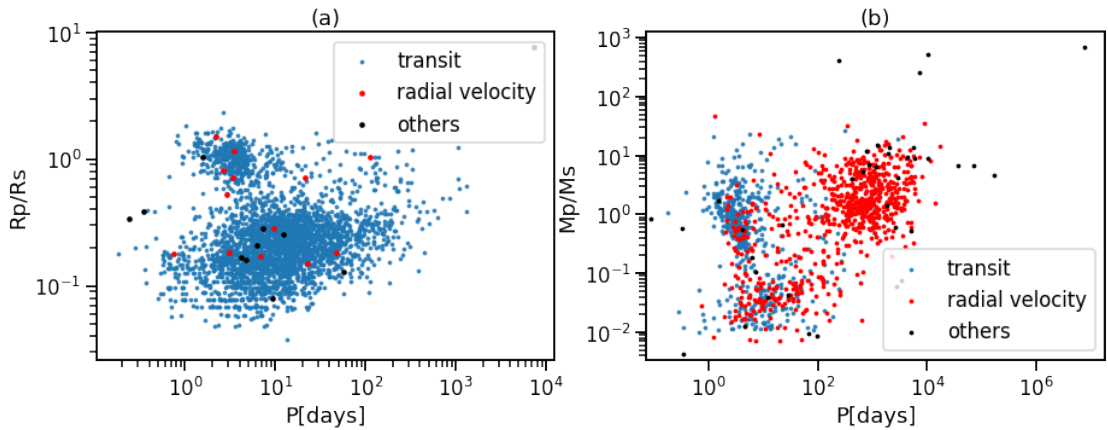

**Figure 3:** The distributions of confirmed exoplanets for all types of host stars in the Archive; (a) in the period-radius plane, and (b) in the mass-radius plane.

As is discussed above, the detection bias leads to different statistics, which means it cannot be ignored or underestimated. So, it is crucial to correct the bias if we want a general conclusion. As a result, such conclusions are always within a given range because of the incompleteness of the data. If the simulation to estimate the yields of exoplanets in the future is based on those general conclusions, the result will be restricted to the range. Here we exploit the dataset of transiting exoplanets to predict the distribution of planetary parameters of new transiting exoplanets in a straightforward manner.

The transiting exoplanets we used in this study are those detected by the Kepler telescope during 4 years (or 1459 days) of its mission from 2009-05-13 00:01:07Z to 2013-05-11 12:16:22Z. The Kepler mission's specified exoplanet detection criterion is a minimum of three observed transit events [*Twicken et al.* 2016]. Thus, most of Kepler's confirmed exoplanets are those with an orbital period less than 486 days. There are some exceptions; for example, the maximum orbital period among the confirmed exoplanets in the Archive, listed under the Kepler Object of Interest (KOI) Catalog Q1–Q17 Data Release 25 (DR 25) [*Twicken et al.* 2016], is Kepler-167e, which has an orbital period of 1071 days. There is no way for Kepler to observe its 3 transits and thus Kepler-167e must be confirmed by other means. Nevertheless, the Kepler data are used to compute the probability density function for this study, since they contain realistic observational information of exoplanets discovered using the transit method. In particular, they also contain the transit probability bias related to the stellar radius and semi-major axis, which will be discussed in the next section.

## 3. Methodology
### 3.1. Sample Selection

The sample we used to obtain the probability density function of the orbital period of exoplanets detected by Kepler is the confirmed exoplanets listed in Kepler Object of Interest (KOI) Catalog Q1–Q17 Data Release 25 (DR 25) [*Twicken et al.* 2016]. The Kepler photometer acquired data at 29.4-minute intervals, known as "long cadences". Science acquisition of Q1 data began at 2009-05-13 00:01:07Z, and acquisition of Q17 data concluded at 2013-05-11 12:16:22Z. This time period is 1459 days and contains 71,427 long cadence intervals. Science data acquisition was interrupted periodically: monthly for data downlink, quarterly for maneuvering to a new roll orientation, and once every three days for reaction wheel desaturation [*Jenkins et al.* 2010]. Since those interruptions are periodic and very short, we ignore them. In addition to the normal interruption, data acquisition was suspended for 11.3 days (555 long-cadence samples) in Q16, and 1145 long-cadence intervals were excluded from searches for transiting planets because of data anomalies. Only if the transit had occurred three times, and one of them exactly occurred in the interrupted period — which means the orbital period is from 364.75 days to 486.3 days — can we miss this exoplanet. Since exoplanets with such long orbital periods cannot be observed by TESS for the required two transits, we do not discuss this effect. We take those interruptions into consideration when calculating the signal-to-noise ratio.

TESS is searching for small transiting planets, which leads primarily to the selection of bright, cool dwarfs [*Stassun et al.* 2018]. The Candidate Target List (CTL) includes both dwarfs and subgiants. In order to match the CTL, we filter the confirmed exoplanets in KOI by excluding those with giant host stars. For the exoplanets that have stellar information of Yerkes spectral classification, we only choose VI (subgiants) and V (dwarfs) stars in our sample, while for those lacking spectral information, we limit the stellar surface gravity (log g) from 3.8 to 5.0 (Figure 4) [*Allen* 1956]. After filters, the size of the final sample is 2141.

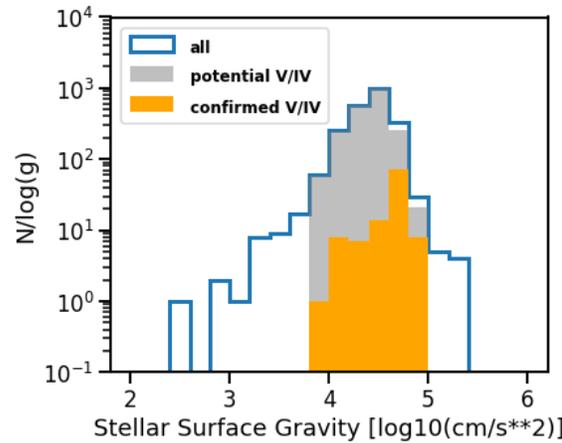

**Figure 4:** The histogram of log g for host stars of all confirmed exoplanets in DR25 (blue), host stars that are confirmed dwarfs and subgiants (orange), and host stars whose surface gravity (log g) from 3.8 to 5.0 (gray).

To match the CTL, we separate the sample into M dwarfs (< 4500K) and AFGK-type stars (> 4500K) according to the stellar effective temperature and adjust the proportion in the later analysis. The size of the subsample of M dwarfs is 218, and that of AFGK-type stars is 2023. The fitting result is shown in Figure 5.

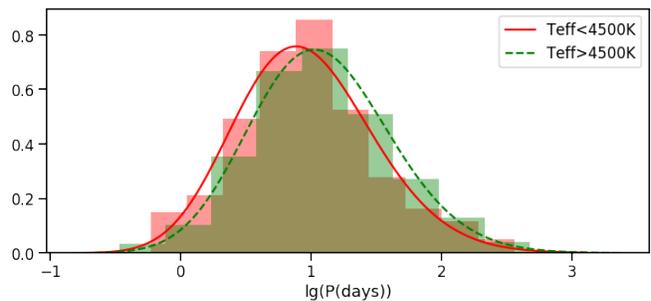

**Figure 5:** The normalized histogram of orbital period for each subsample. The solid and dashed line is the fitting probability density function for subsample 1 and subsample 2, respectively.

### 3.2. Probability density function expression

The purpose of this study is to obtain the orbital period distribution of exoplanets discovered by TESS from Kepler data. Comparison between two theoretical probability density functions of exoplanets' orbital period discovered by TESS and Kepler is made; then the ratio can be calculated based on the different input catalogues and mission parameters. The ratio is a function of orbital period. The probability density function (pdf) of Kepler is obtained by fitting the data (Figure 5), and then we obtain the probability density function of TESS by multiplying the theoretical ratio with the fitting pdf of Kepler.

### 3.2.1. Probability expression

Orbital period is a continuous variable, which means the theoretical probability of a specific orbital period is 0 and only the integral of probability density function makes sense. The probabilities mentioned in the following analysis are all obtained by taking the limit which denotes the values of probability density function at a certain point. Based on Bayes' theorem, the probability that the orbital period of a detected exoplanet is $P$ days is:

$$\text{Prob}(P|TESS) = \text{Prob}(TESS|P) \cdot \text{Prob}(P)/\text{Prob}(TESS) \quad (1)$$

where $TESS$ denotes the event that the exoplanet is detected by TESS and $\text{Prob}(TESS|P)$ can be expanded as:

$$\text{Prob}(TESS|P) = \text{Prob}(tr|P) \cdot \text{Prob}(Ntrs_T|P, tr) \cdot \text{Prob}(SNR_T > SNRT_{min}|P, tr, Ntrs_T) \quad (2)$$

where $\text{Prob}(tr|P)$ is the geometric probability of detecting a transit around a star for a fixed period, $\text{Prob}(Ntrs_T|P, tr)$ is the probability of observing the transit(s) more than $N$ times during the finite observing baseline of observations for TESS for a fixed period, given that the transit is detected, and $\text{Prob}(SNR_T > SNRT_{min}|P, tr, Ntrs_T)$ is the probability that the signal-to-noise ratio (SNR) of the exoplanet is higher than the threshold given that it transits at least $N$ times over the course of the observations.

Then the equation (1) is expanded as:

$$\text{Prob}(P|TESS) = \frac{\text{Prob}(tr|P) \cdot \text{Prob}(Ntrs_T|P, tr) \cdot \text{Prob}(SNR_T > SNRT_{min}|P, tr, Ntrs_T) \cdot \text{Prob}(P)}{\text{Prob}(TESS)} \quad (3)$$

Likewise, repeat the above analysis but for Kepler:

$$\text{Prob}(P|Kepler) = \frac{\text{Prob}(tr|P) \cdot \text{Prob}(Ntrs_K|P, tr) \cdot \text{Prob}(SNR_K > SNRK_{min}|P, tr, Ntrs_K) \cdot \text{Prob}(P)}{\text{Prob}(Kepler)} \quad (4)$$

where $Ntrs_K$ is the minimum required times of transits for Kepler mission, $SNR_K$ is the signal-to-noise ratio detected by Kepler, and $SNRK_{min}$ is the threshold of Kepler pipeline.

### 3.2.2. Calculation of each term

We assume that the planetary radius is much less than the stellar radius ($R_p \ll R_*$) and planetary mass is much smaller than stellar mass ($M_p \ll M_*$). Using the Kepler's third law, the geometric probability is:

$$\text{Prob}(tr|P) = \int \frac{R_*}{a} f_{R_*,a|P}(R_*, a) dR_* da = \int \left(\frac{4\pi^2}{G}\right)^{\frac{1}{3}} R_* M_*^{-\frac{1}{3}} P^{-\frac{2}{3}} f_{R_*,M_*|P}(R_*, M_*) dR_* dM_* \quad (5)$$

where $a$ is the semi-major axis. $f_{R_*,a|P}$ is the joint probability density function of the stellar radius and semi-major axis and $f_{R_*,M_*|P}$ is the joint probability density function of the stellar radius and stellar mass. This probability is only related to the stellar properties except for the orbital period. We assume that the stellar parameters obey the same distribution in each subsample, so the intrinsic stellar parameters of TESS M-dwarf targets are statistically the same as for Kepler M-dwarf targets, even if they differ in number and distance. This integral only contains intrinsic stellar parameters, so the value does not change with the mission. Equation (5) is only the expected value, its uncertainty will be discussed in Section 4.4.

Identifying a transit signal is due to an exoplanet requires a minimum of $N$ times of observed transit events. The probability of an exoplanet with a given orbital period ($P$) can be detected during a given observational duration (t days) can be estimated as:

$$\text{Prob}(Ntrs_T|P, tr) = \begin{cases} 0, & t \leq (N-1)P \\ \frac{t-(N-1) \cdot P}{P}, & (N-1)P < t < NP \\ 1, & t \geq N \cdot P \end{cases} \quad (6)$$

Figure 6 shows the probability that an exoplanet with a given orbital period can be discovered. TESS has different observation duration in different sky-areas due to the multiple sectors, thus the probability varies from one sky-area to another.

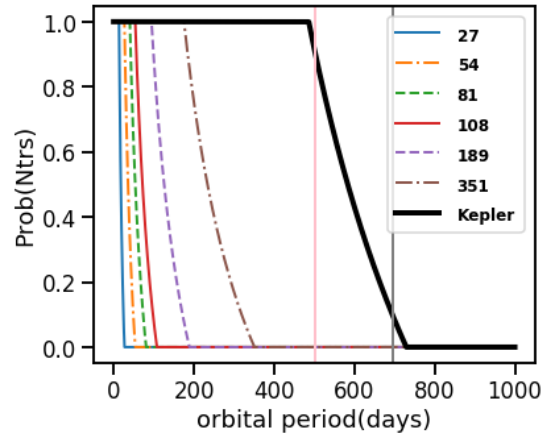

**Figure 6:** Probability of observing two transits (TESS) or three transits (Kepler's detection criterion) during the finite observational duration of 27, 54, 81, 108, 189, or 351 days for TESS, and 1459 days for Kepler. The gray vertical line is 694.76 days beyond which the probability that it can be detected by Kepler is less than 10% and the pink vertical line is 503.10 days within which the probability is higher than 90%.

Here we discuss the impact of the condition on the number of transits observed in the term $\text{Prob}(\text{SNR}_T > \text{SNRT}_{min}|P, tr, Ntrs_T)$. Under the condition where the orbital period ($P$) is fixed, there are three scenarios introduced in equation (6). In the first scenario, the value of $\text{Prob}(Ntrs_T|P, tr)$ is 0, so the value of this term does not matter. In the second scenario, some exoplanets of which the number of times of the transit event is $(N-1)$ will be excluded. Since every exoplanet will be excluded with equal probability, regardless of whether its SNR is higher than the threshold or not, the ratio ($\text{Prob}(\text{SNR}_T > \text{SNRT}_{min})$) will not change. In the third scenario, the probability of that the number of transits observed exceeds $N$ is 100% which means there is no constraint on this term. In conclusion, the condition on the number of transits observed in Prob $(\text{SNR}_T > \text{SNRT}_{min} |P, tr, Ntrs_T)$ has no influence on the final value of the right side of equation (2), so we can remove it. Then the third term in equation (3) is Prob $(\text{SNR}_T > \text{SNRT}_{min} |P, tr)$. Likewise, the third term in equation (4) can be rewritten as $\text{Prob}(\text{SNR}_K > \text{SNRK}_{min}|P, tr)$. The detailed calculation of this term will be discussed in section 3.3.

### 3.2.3. Comparison between two missions

In each subsample, $\text{Prob}(P)$ does not change with the mission. Because the $\text{Prob}(tr|P)$ is a function of stellar properties, the ratio between two of them is not 1 but a distribution. Here we only consider the maximal probabilistic estimations, which means we substitute the random variable with their expected value. According to equation (3) and (4), the probability density function of orbital period of each subsample for the TESS mission can be expressed in terms of that of Kepler:

$$\text{Prob}_i(P|TESS) = c_i \text{Prob}_i(P|Kepler) \frac{\text{Prob}_i(Ntrs_T|P, tr) \cdot \text{Prob}_i(\text{SNR}_T > \text{SNRT}_{min}|P, tr)}{\text{Prob}_i(Ntrs_K|P, tr) \cdot \text{Prob}_i(\text{SNR}_K > \text{SNRK}_{min}|P, tr)}$$
$$i = 1, 2 \qquad (7)$$

where $i = 1$ denotes the subsample of M dwarfs and $i = 2$ denotes the subsample of AFGK-type stars, and $c_i$ is the normalization coefficient. Since $\text{Prob}(Kepler)$ and $\text{Prob}(TESS)$ in equation (3) and (4) are constants, they can be incorporated into the normalization constants.

For Kepler DR25, the required $Ntrs_K$ is 3, and we perform analysis for 2 transits-criterion and 1 transit-criterion for TESS mission.

### 3.3. SNR model

The transit signal-to-noise is $\Delta F_*/\sigma_*$, where $\Delta F_*$ is the change in stellar flux during transit and $\sigma_*$ is the uncertainty in stellar flux during transit.

We exclude grazing transits, for which the planet never completely obscures the star. The signal for a full transit is:

$$\frac{\Delta F_*}{F_*} = \left(\frac{R_p}{R_*}\right)^2 \tag{8}$$

where we have ignored limb-darkening. The Poisson (shot) noise, $\sigma_*$, in the many-photon limit is simply $\sqrt{N}$, where $N$ is the number of photons detected in transit. The precision is the noise divided by the stellar flux, $\sigma_*/F_* = 1/\sqrt{N}$. The signal to noise ratio is therefore:

$$\frac{\Delta F_*}{\sigma_*} = \frac{\Delta F_*/F_*}{\sigma_*/F_*} = \left(\frac{R_p}{R_*}\right)^2 \sqrt{N}. \tag{9}$$

Technically transit photometry is a differential measurement, but we optimistically assume that the uncertainty outside the transit baseline is negligible, which is a reasonable assumption for Kepler or TESS.

Following *Cowan et al.* [2015], the number of in-transit photons detected over the course of a mission is:

$$N = N_{trs} f_t P A \left(\frac{R_*}{r}\right)^2 \int_{\lambda_1}^{\lambda_2} \tau \pi B(\lambda, T_*) \left(\frac{\lambda}{hc}\right) d\lambda, \tag{10}$$

where $N_{trs}$ is the time of transit events during the total observation baseline, $f_t$ is the fraction of time the planet spends in transit, $A$ is the collecting area of the telescope, $\tau$ is the system throughput (photon conversion efficiency = electrons out per photon in), $P$ is the orbital period, $r$ is the distance to the star, $\lambda_1 - \lambda_2$ is the bandpass, and $B(\lambda, T_*)$ is the Planck function of the star.

The fraction of time the planet spends in transit depends on both planetary and stellar parameters. The length of the chord transited by the planet is $l = 2\sqrt{R_*^2 - b^2}$, where $b$ is the planet's impact parameter. The fraction of time spent in transit is therefore the ratio of the chord to the circumference of the orbit:

$$f_t = \frac{2R_*\sqrt{1-b^2}}{2\pi a} = \frac{R_*\sqrt{1-b^2}}{\pi a} \tag{11}$$

where $a$ is the planet's semi-major axis and we have adopted a circular orbit and constant transverse velocity for simplicity. This introduces a nuisance variable, $b$, but we can marginalize over it:

$$<f> = \frac{R_*}{\pi a} \int_0^1 \sqrt{1-b^2} db = \frac{R_*}{4a}. \tag{12}$$

The signal-to-noise ratio is, therefore:

$$\text{SNR} = \frac{R_p^2}{\sqrt{a}} \sqrt{\frac{N_{trs}AP}{4R_*r^2} \int_{\lambda_1}^{\lambda_2} \tau\pi B(\lambda,T_*)\left(\frac{\lambda}{hc}\right) d\lambda}, \tag{13}$$

where we have isolated all of the star- and mission-dependent parameters under the radical.

In order to express the signal-to-noise in terms of the orbital period rather than the semi-major axis, we make use of Kepler's 3rd Law. As the value of SNR is used to calculate the probability, the expected value of $Ntrs$ will be applied but not the single value of every exoplanet. In other words, we are only concerned with the statistic. Then we can plug $<Ntrs> = t_m/P$ into the above equation, of which the error will be discussed in Section 4.4.1:

$$\text{SNR} = R_p^2 P^{-\frac{1}{3}} \left(\frac{4\pi^2}{GM_*}\right)^{\frac{1}{6}} \sqrt{\frac{At_m}{4R_*r^2} \int_{\lambda_1}^{\lambda_2} \tau\pi B(\lambda,T_*)\left(\frac{\lambda}{hc}\right) d\lambda}, \tag{14}$$

We can think of the signal-to-noise as a separable function, but the stellar effective temperature in the Planck function integral makes it impossible to separate the stellar parameters from the mission parameters. Given that we determine the subsample according to effective temperature, we assume that the effective temperature obeys the same distribution for different orbital period bin. So we can use the mean value of it ($<T_*>$) to calculate the $P(SNR > \text{SNR}_{min} | P, tr, Ntrs)$ and then think of the signal-to-noise as a separable function:

$$\begin{aligned} \text{SNR} &= f(\mathbb{P})g(\mathbb{S})h(\mathbb{M},\mathbf{T}_*) \\ f(\mathbb{P}) &= R_p^2 P^{-\frac{1}{3}} \\ g(\mathbb{S}) &= \left(\frac{4\pi^2}{GM_*}\right)^{\frac{1}{6}} \sqrt{\frac{1}{4R_*}} \\ h(\mathbb{M},\mathbf{T}_*) &= \sqrt{\frac{At_m}{r^2} \int_{\lambda_1}^{\lambda_2} \tau\pi B(\lambda,T_*)\left(\frac{\lambda}{hc}\right) d\lambda} \end{aligned} \tag{15}$$

where $\mathbb{P} \equiv \{R_p, P\}$ are the planetary parameters, $\mathbb{S} \equiv \{M_*, R_*\}$ are the stellar parameters, and $\mathbb{M} \equiv \{A, \tau, t_m, \lambda_1, \lambda_2, r\}$ are the mission parameters. Since the distance to the star ($r$) is not the intrinsic property but depends on the input catalogue of the mission, we take it as a mission parameter. The TESS and Kepler mission parameters are $\mathbb{M}_T$ and $\mathbb{M}_K$. So for a given transit event, if it is detected by Kepler and TESS respectively, the SNR should be:

$$\text{SNR}_K = f(\mathbb{P})g(\mathbb{S})h(\mathbb{M}_K,T_*); \quad \text{SNR}_T = f(\mathbb{P})g(\mathbb{S})h(\mathbb{M}_T,T_*) \tag{16}$$

So the $\text{SNR}_T$ can be expressed in terms of $\text{SNR}_k$:

$$\mathrm{SNR}_T = \mathrm{SNR}_K \frac{\mathrm{h}(\mathbb{M}_T, T_*)}{\mathrm{h}(\mathbb{M}_K, T_*)} = k(\mathbb{M}_T, \mathbb{M}_K, T_*) \cdot \mathrm{SNR}_K \qquad (17)$$

where $k$ is the ratio of $\mathrm{h}(\mathbb{M}_T, T_*)/\mathrm{h}(\mathbb{M}_K, T_*)$, which is a function of $\mathbb{M}_T, \mathbb{M}_K$ and $T_*$. So, $k$ is a constant for the certain subsample and certain observing baseline for TESS. The values of mission parameters and the response functions are given on the website of Kepler (https://keplerscience.arc.nasa.gov/the-kepler-space-telescope.html) and TESS (https://heasarc.gsfc.nasa.gov/docs/tess/the-tess-space-telescope.html). We use the mean value of effective temperature in each subsample as $T_*$, the error range will be discussed in 4.5. As $r$ is a random variable, we generate it randomly under its distribution for Kepler and TESS, respectively, to calculate $k$ for 1000 times and take its mean value. The distribution of the distance to CTL stars is given on the portal (https://filtergraph.com/ 1371737). The results of $k$ according to different baselines ($t_m$) for each subsample are listed in Table 1.

**Table 1**

|         | $T_*$   | k (27 days) | k (54 days) | k (81 days) | k (108 days) | k (189 days) | k (351 days) |
|---------|---------|-------------|-------------|-------------|--------------|--------------|--------------|
| < 4500K | 3974.41 | 0.0327      | 0.0463      | 0.0567      | 0.0655       | 0.0866       | 0.1180       |
| > 4500K | 5653.53 | 0.0621      | 0.0878      | 0.1075      | 0.1242       | 0.1643       | 0.2239       |

The probability that the SNR caused by an exoplanet with a given orbital period is higher than the threshold is a complex convolution including the planetary parameters and stellar parameters without the constant mission parameters. As is defined above, in each subsample the stellar parameters and planetary parameters do not change with the mission. Therefore, the complex convolution to calculate the probability is the same for both missions, and the difference between two probabilities is only due to the different mission parameters, which will be explained in more detail below.

Suppose in subsample $i$ ($i = 1,2$) that the probability density functions of SNR caused by the exoplanet with a fixed period for TESS and Kepler, are $f_{\mathrm{SNRT}i}(\mathrm{SNR}|P, tr)$ and $f_{\mathrm{SNRK}i}(\mathrm{SNR}|P, tr)$. For a series of transit events, if they are observed by Kepler, the distribution of SNR will be $f_{\mathrm{SNRK}i}(\mathrm{SNR}|P, tr)$, and if they are observed by TESS, the distribution will be $f_{\mathrm{SNRT}i}(\mathrm{SNR}|P, tr)$. For a certain event, if it is observed by Kepler, the value of SNR is $\mathrm{SNR}_K$, and if it is observed by TESS, the value is $\mathrm{SNR}_T$, which means:

$$f_{\mathrm{SNRT}i}(\mathrm{SNR}_T|P, tr) = f_{\mathrm{SNRT}i}(k \cdot \mathrm{SNR}_K|P, tr) = f_{\mathrm{SNRK}i}(\mathrm{SNR}_K|P, tr) \qquad (18)$$

So the relation between two probability density function is:

$$f_{\text{SNRT}i}(SNR|P,tr) = f_{\text{SNRK}i}(SNR/k|P,tr) \tag{19}$$

And the probability can be calculated as:

$$\text{Prob}_i(\text{SNR}_T > \text{SNRT}_{min}|P,tr)$$

$$= \int_{\text{SNRT}_{min}}^{\infty} f_{\text{SNRT}i}(SNR'|P,tr) dSNR'$$

$$= \int_{\text{SNRT}_{min}}^{\infty} f_{\text{SNRK}i}\left(\frac{SNR'}{k}\Big|P,tr\right) dSNR'$$

$$= \int_{\frac{\text{SNRT}_{min}}{k}}^{\infty} k \cdot f_{\text{SNRK}i}(SNR''|P,tr) dSNR''$$

$$= k \cdot \text{Prob}_i\left(\text{SNR}_K > \frac{\text{SNRT}_{min}}{k}\Big|P,tr\right) \tag{20}$$

where we use the integral substitution. Only for a given stellar sub-sample and a given observing baseline can the substitution work. So the ratio of the probability is:

$$\frac{\text{Prob}_i(\text{SNR}_T > \text{SNRT}_{min}|P,tr)}{\text{Prob}_i(\text{SNR}_K > \text{SNRK}_{min}|P,tr)} = k \cdot \frac{\text{Prob}_i\left(\text{SNR}_K > \frac{\text{SNRT}_{min}}{k}\Big|P,tr\right)}{\text{Prob}_i(\text{SNR}_K > \text{SNRK}_{min}|P,tr)} \tag{21}$$

According to the analysis in Section 3.2.2, adding a condition where at least 3 transits are observed has no influence on the right side of equation (7) for orbital periods within 4 years. So, the ratio of SNR term can be calculated with the following substitution, in order to take use of the Kepler data that is on condition that every detected exoplanet has at least 3 transits.

$$\text{Prob}_i\left(\text{SNR}_K > \frac{\text{SNRT}_{min}}{k}\Big|P,tr\right) \rightarrow \text{Prob}_i\left(\text{SNR}_K > \frac{\text{SNRT}_{min}}{k}\Big|P,tr,3trs_K\right)$$

$$\text{Prob}_i(\text{SNR}_K > \text{SNRK}_{min}|P,tr) \rightarrow \text{Prob}_i(\text{SNR}_K > \text{SNRK}_{min}|P,tr,3trs_K) \tag{22}$$

Then the situation for the Kepler mission is that the probability can be obtained from the known dataset based on the law of large numbers,

$$\frac{\text{Prob}_i\left(\text{SNR}_K > \frac{\text{SNRT}_{min}}{k}\Big|P,tr,3trs_K\right)}{\text{Prob}_i(\text{SNR}_K > \text{SNRK}_{min}|P,tr,3trs_K)} = \frac{N_{Pi}\left(\text{SNR}_K > \frac{\text{SNRT}_{min}}{k}\right)/N_{Pi}}{N_{Pi}(\text{SNR}_K > \text{SNRK}_{min})/N_{Pi}}$$

$$= \frac{N_{Pi}\left(\text{SNR}_K > \frac{\text{SNRT}_{min}}{k}\right)}{N_{Pi}(\text{SNR}_K > \text{SNRK}_{min})} \tag{23}$$

where $N_{pi}$ is the total number of exoplanets with given orbital period ($P$ days), which has been detected at least 3 times transits. For the ones with bracket, only the exoplanet meeting the condition described in the brackets are counted.

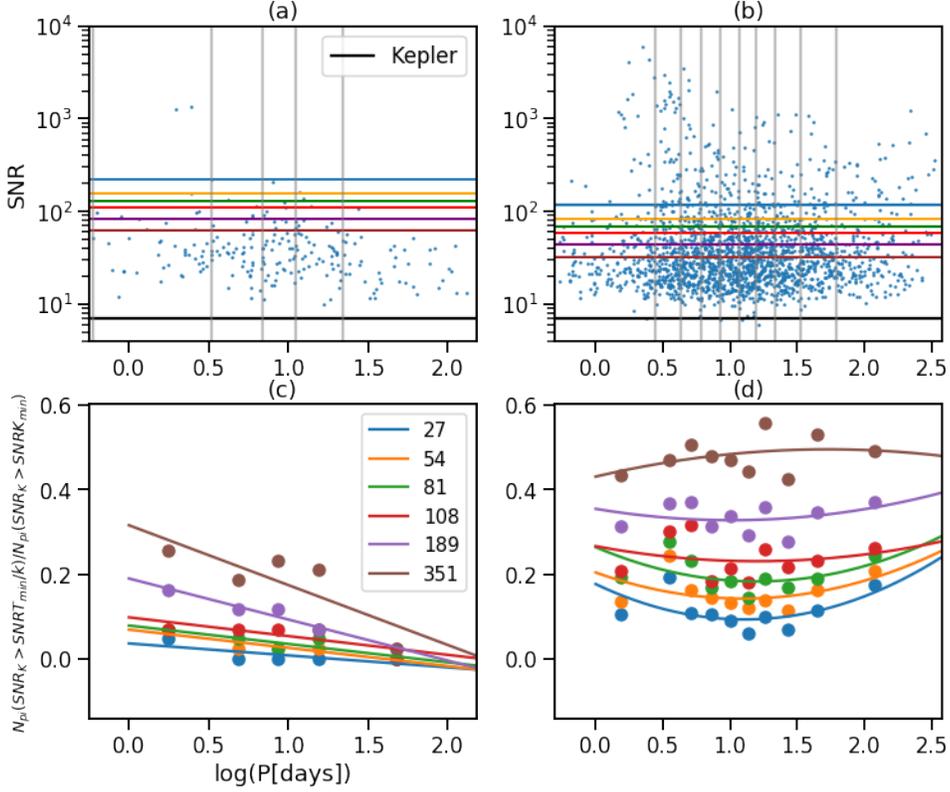

**Figure 7**: ***Top:*** The scatters show the orbital period and SNR of the exoplanets in the sample. The horizontal lines in different colors denote the thresholds for the different observational durations and the legend is shown in (a). (a): sub-sample 1 (<4500K). (b): sub-sample 2 (>4500K). The gray vertical line shows the boundary of each bin. There are 43 exoplanets in each bin in panel (a) and 202 exoplanets in each bin in panel (b). ***Bottom:*** (c) and (d) show the ratio of probability for each bin and result of curve fitting.

The result can derive from the Kepler dataset by dividing the data into different bins of orbital period which contain a certain number of exoplanets (Table 2, Table 3), and then performing linear regression or polynomial curve fitting based on the trend shown in the scatter (Figure 7). For subsample 1 ($T_{eff}$ < 4500K), each bin holds 43 exoplanets, and for subsample 2 ($T_{eff}$ > 4500K), each bin holds 202 exoplanets. We drop the data with the longest orbital period to make sure each bin has the same number of exoplanets. Here, we adopt the SNR threshold of 7.1 that was calculated for Kepler by *Jenkins et al.* [2002] and 7.3 for TESS applied *by Sullivan et al.* [2015]. The results are shown in Table 2 for sub-sample 1 and Table 3 for sub-sample 2. If the fitting value is negative, it is taken as 0. Then we obtain the fitting analytic formula of the ratio of the terms about SNR in equation (7) which can be calculated easily for continuous orbital period.

**Table 2:** The ratio of $(N_P \ (\mathrm{SNR}_K > (\mathrm{SNRT}_{min})/k))/(N_P \ (\mathrm{SNR}_K > \mathrm{SNRK}_{min}))$ in different bins in sub-sample 1 (< 4500 K). There are 43 exoplanets in each bin and 10 bins in total. The mean values of each bin are listed at the top line. The last two columns show the parameters of the fitting curve: 'a' is the coefficient of the result of linear regression, and 'b' is the intercept.

| Mean bin-value | 1.76 | 4.87 | 8.64 | 15.33 | 47.90 | a | b |
|---|---|---|---|---|---|---|---|
| 27 days | 0.05 | 0.00 | 0.00 | 0.00 | 0.00 | − 0.03 | 0.04 |
| 54 days | 0.07 | 0.02 | 0.02 | 0.02 | 0.00 | − 0.04 | 0.07 |
| 89 days | 0.07 | 0.05 | 0.02 | 0.05 | 0.00 | − 0.04 | 0.08 |
| 108 days | 0.07 | 0.07 | 0.07 | 0.07 | 0.00 | − 0.04 | 0.10 |
| 189 days | 0.16 | 0.12 | 0.12 | 0.07 | 0.02 | − 0.10 | 0.19 |
| 351 days | 0.26 | 0.19 | 0.23 | 0.21 | 0.02 | − 0.14 | 0.32 |

**Table 3:** The ratio of $(N_P \ (\mathrm{SNR}_K > (\mathrm{SNRT}_{min})/k))/(N_P \ (\mathrm{SNR}_K > \mathrm{SNRK}_{min}))$ in different bins in sub-sample 2 (> 4500 K). There are 202 exoplanets in each bin and 20 bins in total. The mean values of each bin are listed at the top line. The last three columns show the parameters of the fitting curve: $y = k_1 x^2 + k_2 x + k_3$

| Mean bin-value | 1.55 | 3.52 | 5.10 | 7.21 | 10.06 | 13.57 | 18.07 | 26.69 | 44.42 | 119.98 | $k_1$ | $k_2$ | $k_3$ |
|---|---|---|---|---|---|---|---|---|---|---|---|---|---|
| 27 days | 0.10 | 0.19 | 0.11 | 0.10 | 0.09 | 0.06 | 0.10 | 0.07 | 0.11 | 0.17 | 0.07 | − 0.15 | 0.18 |
| 54 days | 0.13 | 0.24 | 0.16 | 0.14 | 0.13 | 0.12 | 0.14 | 0.11 | 0.16 | 0.21 | 0.05 | − 0.11 | 0.20 |
| 89 days | 0.19 | 0.28 | 0.23 | 0.17 | 0.18 | 0.14 | 0.19 | 0.17 | 0.19 | 0.24 | 0.06 | − 0.14 | 0.26 |
| 108 days | 0.21 | 0.30 | 0.32 | 0.18 | 0.21 | 0.18 | 0.26 | 0.22 | 0.23 | 0.26 | 0.02 | − 0.06 | 0.27 |
| 189 days | 0.31 | 0.37 | 0.37 | 0.31 | 0.34 | 0.29 | 0.36 | 0.28 | 0.35 | 0.37 | 0.03 | − 0.05 | 0.35 |
| 351 days | 0.43 | 0.47 | 0.50 | 0.48 | 0.47 | 0.44 | 0.56 | 0.43 | 0.53 | 0.49 | − 0.02 | 0.07 | 0.43 |

## 4. Results and Discussion

### 4.1. The results of 2-transit and 1-transit criterion for M-Dwarfs and Sun-like stars

Based on equation (7), we obtain the predicted distribution of orbital periods for TESS (Figure 8). For the 2-transit criterion, the result is Figure 8 (a) for sub-sample 1 (< 4500K) and (b) for sub-sample 2. For the one-transit criterion, the result is shown in Figures 8 (c) and (d).

As is shown in Figure 8, for exoplanets orbiting M dwarfs, the orbital period tends to be shorter compared to those with AFGK-type stars, and the peak value becomes shorter with the reduction of the observing baseline, while the trend is not that obvious for sub-sample 2. It indicates that the M dwarfs tend to have inner short-period exoplanets detected. But it also may result from the lack of data to estimate the probability of SNR in sub-sample 1.

For exoplanets orbiting stars with $T_{eff} > 4500$ K, when requiring 2 transits, the orbital period distribution is much more sensitive to the observing baseline than when requiring at least one transit, especially for longer orbital periods. There is an obvious cut-off at the longer orbital period for sub-sample 2 when requiring 2 transits.

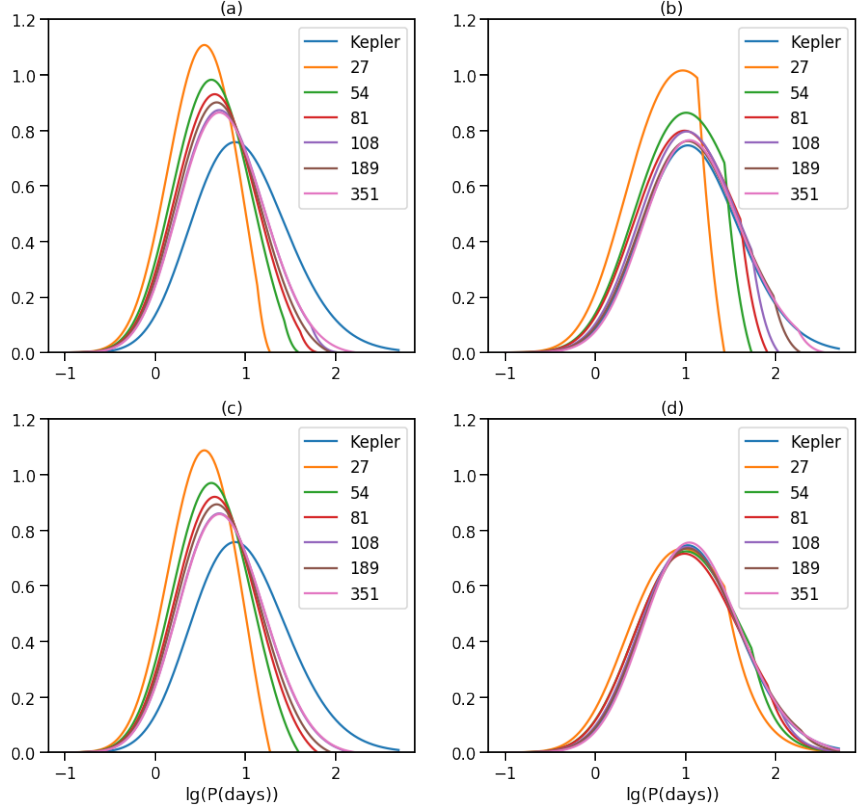

**Figure 8**: Top: The probability density function of orbital period for different observing baselines of the TESS mission if at least 2 transits are needed: (a) the effective temperature ($T_{eff}$) of the host star is < 4500 K, and (b) the $T_{eff}$ of the host star is > 4500 K. Bottom: The probability density function of orbital period for different observing baselines of the TESS mission if only one transit is needed: (c) the $T_{eff}$ of the host star is < 4500 K, and (d) the $T_{eff}$ of the host star is > 4500 K.

**Table 4:** Mean orbital period (MP) and detection range within 1σ of TESS's most frequently detected orbital period of exoplanets in each duration of observations for each sub-sample, assuming the detection criteria is 2 transits and one transit. The Kepler's most frequently detected exoplanet orbital period is also shown under Kepler's detection criteria of 3 transits.

| Duration of Observation (days) | 27 | 54 | 81 | 108 | 189 | 351 | Kepler |
|---|---|---|---|---|---|---|---|
| MP(days) - sub1 -2 transits | 3.16 | 4.10 | 4.67 | 5.52 | 5.07 | 5.67 | 9.27 |
| 1σ (days) - sub1 -2 transits | 1.44-6.93 | 1.69-9.95 | 1.82-12.00 | 2.01-15.19 | 1.89-13.56 | 2.01-16.03 | 2.73-31.52 |
| MP(days) - sub2 -2 transits | 5.17 | 7.57 | 8.59 | 9.71 | 11.05 | 12.16 | - |
| 1σ (days) - sub2 -2 transits | 2.20-12.14 | 2.89-19.83 | 3.05-24.23 | 3.38-27.95 | 3.51-34.29 | 3.80-38.97 | - |
| MP(days) - sub1 -1 transit | 3.12 | 4.05 | 4.62 | 5.56 | 5.00 | 5.60 | 12.23 |
| 1σ (days) - sub1 -1 transit | 1.41-6.93 | 1.65-9.95 | 1.78-12.01 | 1.96-15.73 | 1.86-13.47 | 1.97-15.93 | 3.54-42.20 |
| MP(days) - sub2 -1 transit | 8.56 | 10.44 | 10.75 | 11.38 | 12.02 | 12.49 | - |
| 1σ (days) - sub2 -1 transit | 2.68-27.33 | 3.18-34.26 | 3.18-36.37 | 3.43-37.75 | 3.53-40.89 | 3.75-41.65 | - |

Although the orbital period is unconstrained when requiring only 1 transit, the panel (c) shows the impossibility of detecting long-period exoplanets orbiting M dwarfs, which is similar to the result when requiring two transits. It is because SNR dominates the probability to detect the long-period but not the criteria of the times of detected transits, for M dwarfs, which is shown in Figure

(6) and Figure (7). One reason is that M dwarfs are so faint that it is hard to confirm the single transit due to the low SNR. The low SNR for M dwarfs also indicates the low occurrence of long-period giant exoplanets that are giant enough to produce the high SNR. In principle, the incompleteness of Kepler data could lead to this result. But according to our introduction, it will not have a significant impact.

### 4.2. Combination of 2 sub-samples

We adopt the result of *Barclay et al.* [2018] that TESS will observe 371($n_1$) M dwarfs hosting exoplanets and 922($n_2$) AFGK-type stars hosting exoplanets. And we assume that the ratio of $n_1$ to $n_2$ is fixed for arbitrary sky-areas. So, the probability density function for the full sample is:

$$\text{Prob}(P|TESS) = c_T(n_1 \text{Prob}_1(P|TESS) + n_2 \text{Prob}_2(P|TESS)) \quad (24)$$

where $c_T$ is a normalization coefficient. The probability density function curve for two-transit criterion and 1-transit criterion are shown in Figure 9.

According to Figure 9, the orbital period distribution for TESS has an obvious cut-off at the long orbital period when at least 2 transits are required, especially for the short observing baseline, which indicates that there is a large incompleteness for long orbital period exoplanets, while there is no big difference between different observing baselines when requiring at least 1 transit.

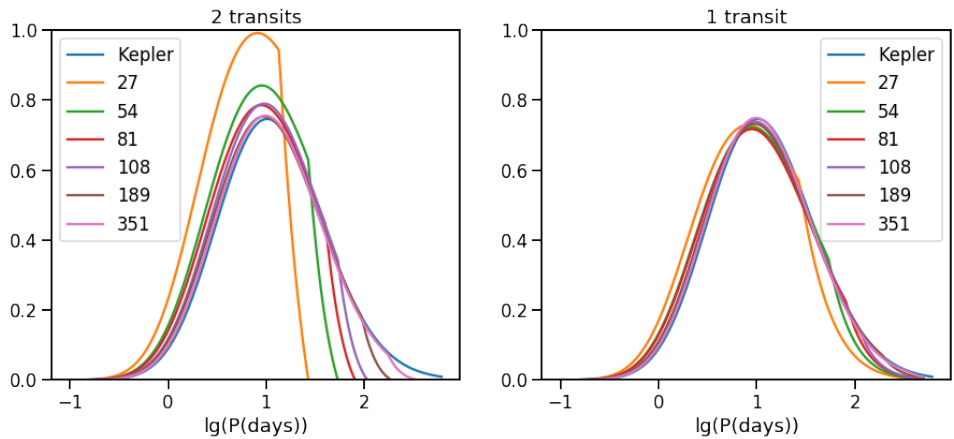

**Figure 9:** Left: The criterion is 2 transits. Right: The criterion is 1 transit.

**Table 5:** Mean orbital period (MP) and detection range within 1σ of TESS's most frequently detected orbital period of exoplanets in each duration of observations, assuming the detection criteria is 2 transits and 1 transit. For comparison, Kepler's most frequently detected exoplanet orbital period is also shown under Kepler's detection criterion of 3 transits.

| Duration of Observation (days) | 27 | 54 | 81 | 108 | 189 | 351 | Kepler |
|---|---|---|---|---|---|---|---|
| MP(days) -2 transits | 5.01 | 7.06 | 8.00 | 9.21 | 10.07 | 10.93 | 11.89 |
| 1σ (days) -2 transits | 2.12-11.82 | 2.64-18.91 | 2.77-23.10 | 3.17-26.76 | 3.19-31.81 | 3.35-35.65 | 3.45-41.04 |
| MP(days) -1 transit | 8.17 | 9.74 | 10.12 | 10.96 | 10.96 | 11.25 | - |
| 1σ (days) -1 transit | 2.55-26.21 | 2.95-32.19 | 2.98-34.31 | 3.21-35.62 | 3.20-37.57 | 3.34-37.91 | - |

## 4.3. Combination of sectors

According to *Barclay et al.* [2018], the number of detected exoplanets of different observing baselines is 543, 273, 101, 193, 108, and 75 for 27 days, 54 days, 81 days, 108-297 days, 324 days, and 351 days. Taking them as the coefficient of each probability density function for a given observing baseline and combining them together, the result is shown in Figure 10 and Table 6.

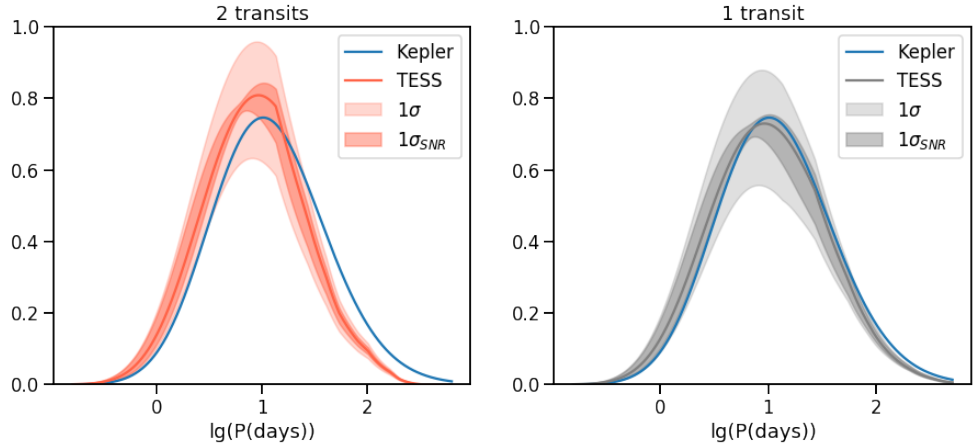

**Figure 10:** Left: The criterion is 2 transits. Right: The criterion is 1 transit. The light shadow is the possible range within 1 standard deviation and the deep shadow is the possible range only considering the uncertainty caused by the term of SNR, which is discussed in Section 4.4.

**Table 6:** Mean orbital period (MP) and detection range within $1\sigma$ of TESS's most frequently detected orbital period of exoplanets, assuming the detection criterion is 2 transits and 1 transit. For comparison, Kepler's most frequently detected exoplanet orbital period is also shown under Kepler's detection criterion of 3 transits.

| Duration of Observation (days) | TESS | Kepler |
|---|---|---|
| MP(days) -2 transits | 8.47 | 11.89 |
| $1\sigma$ (days) -2 transits | 2.75-26.12 | 3.45-41.04 |
| MP(days) -1 transit | 10.09 | - |
| $1\sigma$ (days) -1 transit | 2.99-34.08 | - |

For the two-transits criterion, the distribution has a cut-off at the orbital period longer than 20 days. This is due to the fact that exoplanets with an orbital period longer than 27 days cannot be observed twice in most of the sky-area. For 1-transit criterion, the orbital period distribution is similar to that of Kepler, though it shifts left a little.

## 4.4. Uncertainty analysis
### 4.4.1. Uncertainty of approximating $Ntrs$

To evaluate the influence of approximating $Ntrs$, we generate 1000 mock data to analyze it. The distributions of every parameter contributed to SNR are shown in the first row in Figure 11.

At first, we simulated the SNR when those mock exoplanets detected by Kepler using the precise formula:

$$\text{SNR} = R_p^2 \left(\frac{4\pi^2 P}{GM_*}\right)^{\frac{1}{6}} \sqrt{\frac{N_{trs}A}{4R_*r^2} \int_{\lambda_1}^{\lambda_2} \tau \pi B(\lambda, T_*) \left(\frac{\lambda}{hc}\right) d\lambda}, \quad (25)$$

where $N_{trs}$ is the time of detected transit events. Unlike equation (14) using expected value of all exoplanets with the same orbital period, equation (25) applies to single detected exoplanet. The simulated signals are shown in the lower left panel in Figure 11, and $N_{trs}$ is generated randomly under the Bernoulli distribution.

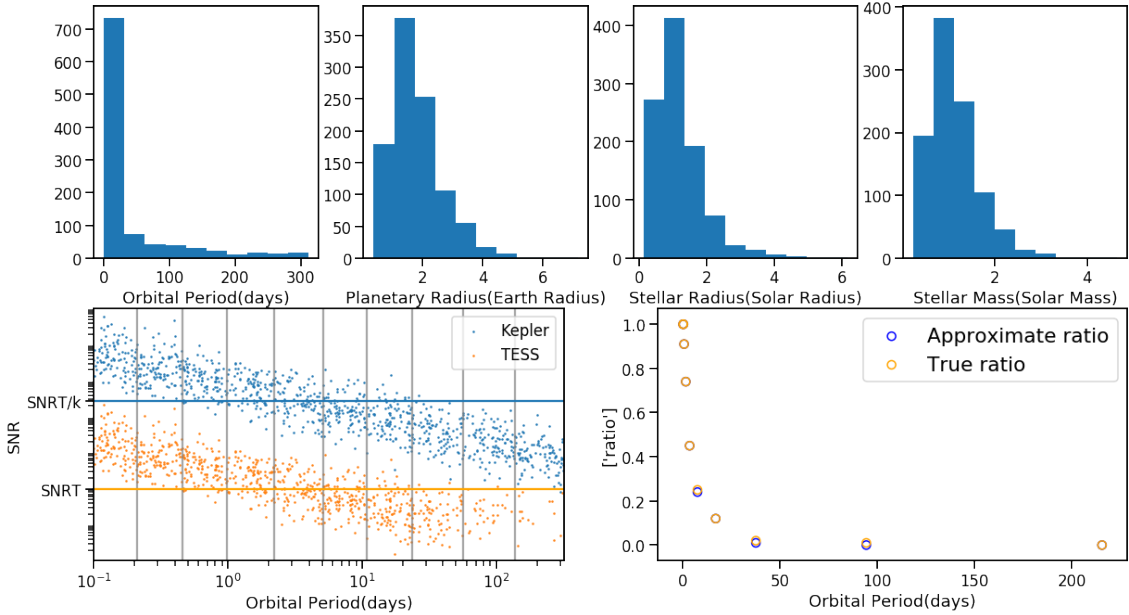

**Figure 11:** Top: The distributions of all parameters contributed to SNR. Bottom: Left: The blue dots represent the signals detected by Kepler, and the orange dots represent the signals detected by TESS. The orange horizontal line is the threshold of TESS mission (SNRT), and the blue horizontal line is SNRT/k introduced in equation (20) which is used to estimate the ratio of the terms about SNR in equation (7). Right: The blue circles represent the ratio estimated based on Kepler data, and the orange circles represent the true ratio derived from the precise SNR of TESS. Some orange circles cover the blue circles as they have the same value.

Then, we apply the same method introduced in Section 3.3 to estimate the ratio of the probability terms about SNR. The substitution of threshold (SNRT/*k*) is shown as blue horizontal line in the lower left panel in Figure 11. Only exoplanets of which the signals are above the threshold can be detectable by TESS. And we use this ratio of the number of exoplanets above the threshold to the number of exoplanets below the threshold to estimate the ratio of the probability terms about SNR for TESS, which is shown in the lower right panel in Figure 11.

In order to compare the estimate to the true ratio, we also simulated the SNR when those mock exoplanets detected by TESS in the 27 days baseline using equation (25), through which $N_{trs}$ is generated randomly under the Bernoulli distribution. The results are shown as orange dots in the lower left panel in Figure 11 and the threshold (SNRT) is the value of the orange horizontal line. The results of true value are shown in the lower right panel of Figure 11 (orange circles). As we can see, the error of our estimation is small, which indicates its validity.

### 4.4.2. Discussion about different detection models

In our model introduced in Section 3, the SNR includes all transit events during the observation. The exoplanet can be confirmed as long as its stacked SNR is above the threshold. In fact, the first single transit event also needs to exceed a threshold so that we can confirm it with following multiple signals. Therefore, some exoplanets with lower single SNR would be excluded in this way, though its total SNR is higher than the threshold due to the multiple stacking. However, those exoplanets will be excluded evenly regardless of their orbital period, so we ignore this threshold in this paper.

### 4.4.3. Uncertainty of Stellar Parameters

As the stellar parameters obey the same distributions for a given subsample, $\text{Prob}(tr|P)$ should not change with the missions as equation (5) indicates. But the integral in equation (5) is only the expected value. In fact, this term is a variable drawn from a random distribution. Although the terms of two missions are drawn from the same distribution, their values may differ, leading to the dispersion of the ratio of the terms for two missions, which means that the ratio is not 1. Here we use different subscribe to distinguish these two variables. If so, equation (7) should be rewritten as follow, and then the uncertainty of the probability density can be obtained by the propagation of uncertainty.

$$\text{Prob}_i(P|TESS)$$
$$= c_i \text{Prob}_i(P|Kepler) \cdot \frac{\text{Prob}(tr|P)}{\text{Prob}(tr|P)} \cdot \frac{\text{Prob}_i(Ntrs_T|P,tr)}{\text{Prob}_i(Ntrs_K|P,tr)}$$
$$\cdot \frac{\text{Prob}_i(SNR_T > SNRT_{min}|P,tr)}{\text{Prob}_i(SNR_K > SNRK_{min}|P,tr)}, i = 1,2 \qquad (26)$$

where the second term (geometric probability term) and the fourth term (SNR term) include the random parameters.

According to equation (5), the uncertainty of geometric probability should be propagated by the uncertainties of the stellar parameters $(R_*, M_*)$. Here, we regard this term as a single variable and obtain the statistics directly (Figure 12, Table 7). According to the propagation of uncertainty, the uncertainty of the ratio is:

$$\sigma_{\mathbf{Prob}(tr|P)/\mathbf{Prob}(tr|P)} = \sqrt{2 \cdot (\sigma_{\mathbf{Prob}(tr|P)}/\mathrm{Prob}(tr|P))^2} \qquad (27)$$

The values are shown in Table 7.

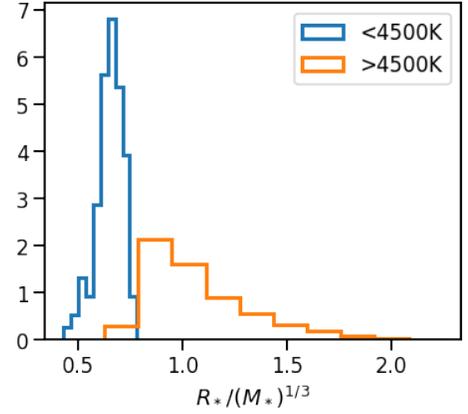

**Figure 12:** The distribution of the random variable term in $\mathrm{Prob}(tr|P)$. $\mathrm{Prob}\,(tr|P) = (4\pi^2/G)^{1/3} R_* M_*^{-1/3} P^{-2/3}$ We plot the histogram of $R_* M_*^{-1/3}$ to illustrate its distribution and obtain its standard deviation to calculate the uncertainty of the ratio by the propagation of uncertainty (equation (27)).

**Table 7:** The standard deviation of random variable terms.

| Statistic | <4500K | >4500K |
|---|---|---|
| $\dfrac{\sigma_{\mathbf{Prob}(tr|P)}}{\mathbf{Prob}(tr|P)}$ | 0.0648 | 0.2571 |
| $\dfrac{\sigma_{\mathbf{Prob}(tr|P)/\mathbf{Prob}(tr|P)}}{\mathbf{Prob}(tr|P)/\mathbf{Prob}(tr|P)}$ | 0.1407 | 0.3380 |

### 4.4.4. Uncertainty of SNR model

As for the uncertainty of the SNR term, the standard deviation of k, or $\sigma_k$, could only be derived from the deviations of effective temperature and distance to the stars. The relative error of $k$ caused by the variance of effective temperature is:

$$\frac{k_{max}(\mathbb{M}_T, \mathbb{M}_K, T') - k_{min}(\mathbb{M}_T, \mathbb{M}_K, T'')}{k(\mathbb{M}_T, \mathbb{M}_K, T_*)} \qquad (28)$$

where $T'$ is the effective temperature that maximizes the $k$ and $T''$ minimizes $k$. For example, the relative error of $k$ for 27-days observation is $1.5 \times 10^{-6}$ for M-dwarfs with effective temperature ranging from 1K to 4500K and $5.2 \times 10^{-9}$ for sun-like stars with effective temperature ranging from 4500K to 10000K, respectively. And the relative errors for other observational baselines are of the same order of magnitude for each subsample. These variances lead to little influence on the fitting results shown in Figure 7, so we ignore the uncertainty of the final result caused by them.

As Section 3.3 introduced, in order to deal with the uncertainty caused by the distance, we generate *r* randomly under its distribution for Kepler and TESS, respectively, to calculate k and take the following steps using it for 1000 times. The 1-sigma range shows in Figure 10 as deep shadows for 2-transit criterion and 1-transit criterion, respectively.

Taking all kinds of uncertainties into consideration, the uncertainty of the probability density function can be obtained by the propagation of the above uncertainties, and the results are shown in Figure 10 as light shadows.

### 4.5. Comparison with previous studies

In order to compare our results with previous studies in the literature, we generated 1000 simulated data according to the given probability density function above by rejection sampling, which is shown in Figure 13.

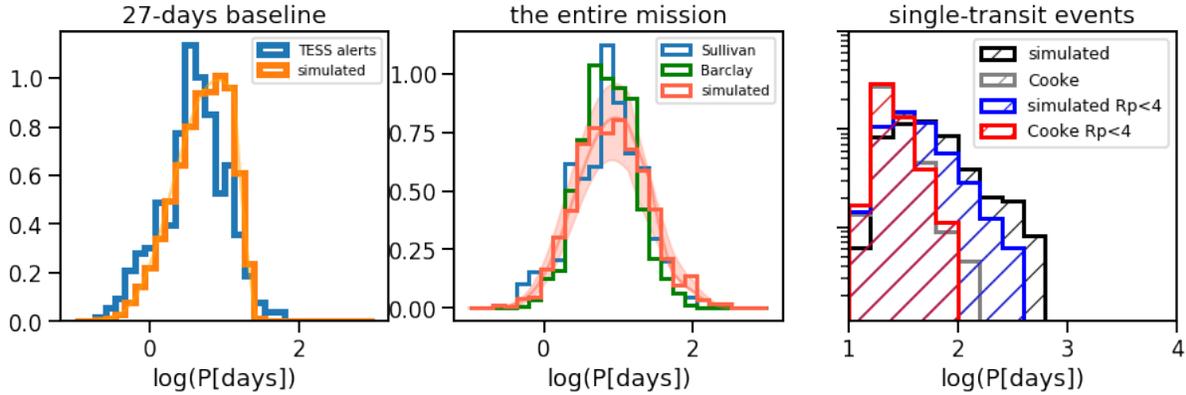

**Figure 13:** Left panel: The blue profile is the histogram of TESS alerts and the orange profile is the histogram of the simulated data under the probability density function of orbital period for 27 days observing baseline. Median panel: The blue profile is the histogram of simulated data by *Sullivan et al.* [2015] and the green profile is the histogram of the simulated data by *Barclay et al.* [2018] and the pink profile is the histogram of simulated data under the probability density function of orbital period for the entire TESS mission. The shadow is the possible range within 1 standard deviation. Right: The orbital period histogram of simulated single transit events is in black. The histogram of single transit events of Sub-Neptune is in blue and red, representing our results and the results of *Cooke et al.* [2018]. The length and range of the bins are the same as those of Fig. 1 from *Villanueva et al.* [2018].

For the two-transits criterion, we compare our results to the newly available TESS alerts data (https://tess.mit.edu/alerts/) and the results from *Sullivan et al*. [2015] and *Barclay et al*. [2018]. Since the available TESS data are only from two sectors, we simulated data according to the probability density function of an orbital period for a 27-day observing baseline. Compared to the TESS alerts, which are from preliminary data and likely to contain false positives, our simulation results show good agreement with the TESS preliminary data. However, our simulation shows that TESS will detect slightly more longer-period exoplanets, and the peak is also slightly wider than that of alerts.

Our simulated sample also contains a larger proportion of longer-period exoplanets compared to both *Sullivan et al.* [2015] and *Barclay et al.* [2018]. The cause might be that the occurrence rates from both *Fressin et al.* [2013] and *Dressing & Charbonneau* [2015], which were used in the studies of Sullivan et al. and Barclay et al., are limited in orbital period to 0.5-85 and 0.5-200 days, respectively. For the longer observing baseline, their input data is not complete.

For the one-transit criterion, we extract the single-transit events from the entire yields shown in Figure 10 and generate 241 data of which histogram in Figure 13, in order to compare it to the result shown in Figure 1 from *Villanueva et al.* [2018]. They show the same trend, though the peak is not as obvious as the result of *Villanueva et al.* [2018]. They assumed that the occurrence of the longer orbital period is equal to the occurrence of the longest complete orbital period. The planet occurrence rates are only complete to periods of 60 days for AFGK-type stars and 30 days for M dwarfs because they derive from planetary radius and orbital period. In this study, we focus on the orbital period, so the distribution of orbital period up to 695 days can derive from the Kepler dataset and the distribution of Kepler shows the low occurrence of the longer orbital period. The results of *Cooke et al.* [2018] are far narrow than both results discussed above. *Cooke et al.* [2018] and *Villanueva et al.* [2018] applied the same input catalogue and used the same occurrence rate, but with different approaches. It can be seen that the detection model of *Cooke et al.* [2018] is more stringent. In this study, we apply the SNR threshold of 7.3 for both multi-transits and single transit, although it may be optimistic.

*Cooke et al.* [2018] also gave the predicted orbital distribution of sub-Neptune detected by TESS, which is listed in the Table 4 from *Cooke et al.* [2018]. In order to compare our results to that, we resample the Kepler data by excluding those exoplanets with radius greater than 4 earth-radius and repeat the above steps introduced in Section 3 to obtain the probability density function of orbital period of sub-Neptune. And then we generate 241 data and plot the histogram for comparison (Figure 13). The lower yields of longer-orbital period sub-Neptune indicates their lower occurrence rate. This trend can also be seen in the results of *Cooke et al*, while the range of our results is wider, which has been discussed above.

## 5. Summary and discussion

The orbital period detectable by TESS varies in different regions of the celestial sphere, due to different observation durations. Transit surveys are strongly biased toward short periods, especially for TESS whose observation period is so short in most of the celestial sphere. Our mathematical analysis finds (Table 5) that, for the two-transit criterion, the expected mean value of most frequent detected orbital period is 5.01 days with a most detected range of 2.12 to 11.82 days in the region with the observation of 27 days. Near the poles where the observational duration is 351 days, the expected mean orbital period is 10.93 days with a most detected range from 3.35 to 35.65 days. For one-transit, the most frequent detected orbital period is 8.17 days in the region with the observation of 27 days and 11.25 days near the poles. These results are illustratively summarized in Figure 14. For the entire TESS mission containing several sectors (Table 6), the mean value of orbital period is 8.47 days for two-transit and 10.09 days for one-transit, respectively. With such orbital periods, the majority of exoplanets detected by TESS will be closer to their host stars than Mercury to our Sun, whose orbital period is 88 days.

The major difference from this study to the previous study is leveraging some of the similarities in biases between TESS and Kepler to cut out many of steps taken to make a mock catalogue. Admittedly, this study involves some approximations, including treating each of the stellar sub-samples as homogeneous and the simplification of the SNR model.

Unlike the previous studies that simulate the detected exoplanet with several properties, we only focus on the orbital period. Previous simulations [e.g., *Barclay et al.* 2018, *Sullivan et al.* 2015, *Huang et al.* 2018, *Cooke et al.* 2018 and *Villanueva et al.* 2018] require the general conclusion of the occurrence of the exoplanets. To do that, they might need to know about the distribution of different properties and the correlation among them. Actually, it is incredibly hard to recover the data due to selection bias, let alone to obtain the general conclusion. Therefore, they have to ignore some relations and simplify the occurrence model of exoplanets to perform the simulation by generating some parameters from a random uniform distribution. If we perform the estimate directly on the data, we can take those parameters and their relation into consideration.

We could further predict the distribution of other planetary properties. For example, the planetary radius affects the SNR, so we could estimate the radius distribution by dividing the radius sample into several subsamples based on orbital period and then perform an analysis on how the SNR makes a difference in the distribution of an individual radius. This could help us to predict the yields of habitable exoplanets.

The significance of this work is not only in its ability to predict but also as a tool to perform analysis between two datasets. For example, we could predict the planetary radius based on the prediction of orbital period distribution that agrees with the real data. If the result is far different from the real yields, this would suggest that there is no dependence between radius and orbital period.

The major reason why we choose to predict the orbital period distribution rather than other properties is that the major difference between TESS and Kepler is the duration of observation. It makes the biggest difference in orbital period, which means the changes in the distribution of orbital period between two missions should be the most notable, making it easier to do follow-on research.

Furthermore, in this study, a vital assumption is that the stellar parameters and planetary parameters which determine the occurrence of exoplanets obey the same distribution for M dwarfs and AFGK –type stars, respectively; this similarity is often applied in the literature. If the yield of TESS is very different from our prediction, the assumption should be questioned. If so, we could change the assumption and perform analysis to see which parameters can dominate the exoplanet occurrence based on the methodology of this study.

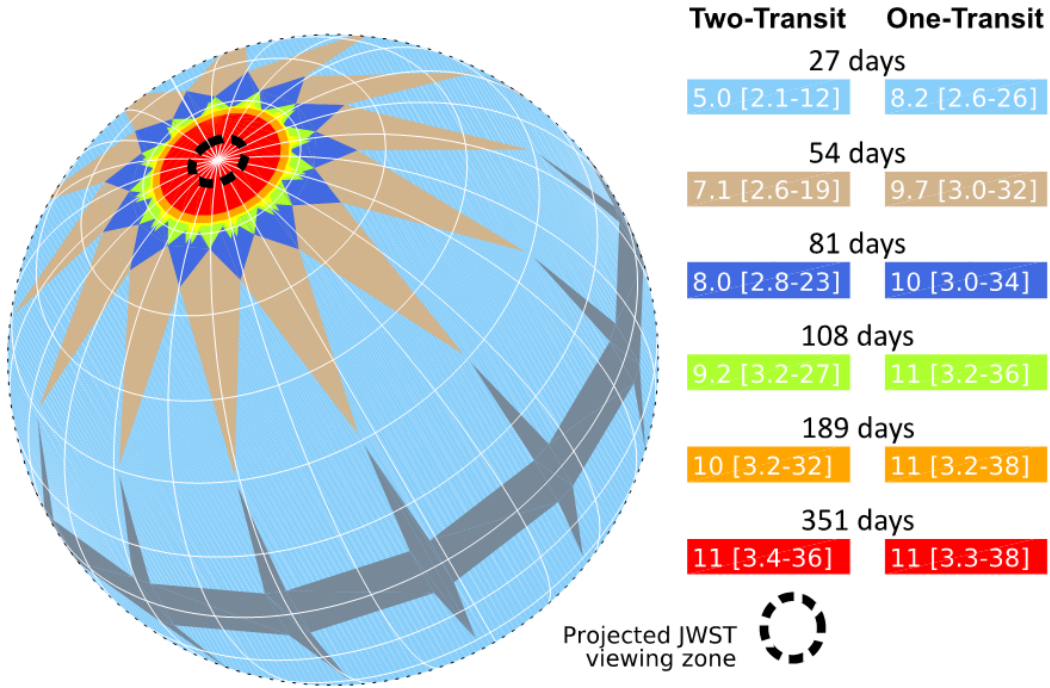

**Figure 14:** Duration of observations on the celestial sphere by TESS, distinguished by color, reproduced after *Ricker et al.* [2015]. Mean orbital period and detection range of TESS's most frequently detected orbital period of exoplanets for two-transit and one-transit criteria are listed by the text inside the color-bars.

**Acknowledgement**

This work was supported by an exoplanet study initiative at the Jet Propulsion Laboratory, California Institute of Technology, under contract with NASA. Author Xuan Ji thanks the Joint Institute for Regional Earth System Science and Engineering, University of California, Los Angeles, for hosting her summer research via a student exchange program. We thank Junyang Long of Chinese University of Hong Kong for helpful comments on the statistical method. Funding for the TESS mission is provided by NASA's Science Mission Directorate and we acknowledge the use of TESS Alert data, which is currently in a beta test phase, from pipelines at the TESS Science Office and at the TESS Science Processing Operations Center.

**Data availability:** The exoplanet data used for this study can be downloaded at the NASA Exoplanet Archive (https://exoplanetarchive.ipac.caltech.edu/) and the TESS Alerts (https://tess.mit.edu/alerts/). The simulated data generated as the results of this study will be available for two years and can be downloaded from the 3rd party repository at https://zenodo.org/record/3245557. Please contact the corresponding author for any questions at Jonathan.H.Jiang@jpl.nasa.gov.